# Interacting antiferromagnetic and ferroelectric domain structures of multiferroics

**Z.V. Gareeva[*,1,2], A.K. Zvezdin[1,3]**

[1] A.M. Prokhorov General Physics Institute, Russian Academy of Sciences, Vavilova St. 38, Moscow, 119991, Russia
[2] Institute of Molecular and Crystal Physics Russian Academy of Sciences, prospect Octyabrya 151, Ufa, 450075, Russia
[3] Fondazione ISI 10133, Torino, Italy

[*] Corresponding author: e-mail gzv@anrb.ru, tel.:+7(3472)313538, fax: +7(3472)311995  zvezdin@gmail.com

Abstract. The correlation between antiferromagnetic and ferroelectric domain structures in multiferroics has been studied. The role of magnetoelectric interactions in a formation of antiferromagnetic domain structure has been analysed. It has been shown the major physical mechanism binding antiferromagnetic domains to ferroelectrics ones is inhomogeneous flexomagnetoelectric interaction of the $P_z\left[(l\nabla)l_z - l_z(\nabla l)\right]$ type. The dependences illustrating rearrangement of antiferromagnetic domain pattern together with change of ferroelectric domains thickness has been performed.

**Introduction** Multiferroics, being materials where ferroelectric and magnetic orders coexist, are currently attracting considerable attention. With an expanding demand for spintronics, nonvolatile ferromagnetic or ferroelectric memory devices, data storage media etc. the interest to investigation of domain patterning in multiferroics is permanently arising. The possible coupling between ferroelectric and antiferromagnetic domains has been discussed in a series of experimental and theoretical works [1 - 14]. The physical mechanisms leading to antiferromagnetic and ferroelectric domains clamping are under discussion.

In this letter, we report the multiferroic antiferromagnetic domain structure and the physical mechanisms responsible for its formation. Two magnetoelectric interactions such as homogeneous Dzyaloshinskii – Moria interaction and inhomogeneous flexomagnetoelectric interaction resulting in magnetic and electric order parameters coupling are known, see for details [15, 16]. We show that the antiferromagnetic domain structure is pinned to ferroelectric domain structure, and that the cornerstone mechanism responsible for this process is inhomogeneous interaction (in other words, flexomagnetoelectric interaction [17]). We study transformation of antiferromagnetic vector distribution induced by the ferroelectric domain structure.

**Model** Let us consider $BiFeO_3$ as a model object for the analysis. The latest experiments have shown the $BiFeO_3$ films demonstrate dramatically high values of electric polarization up to $10^{-4}$ C/cm$^2$ [4]. The domains in multiferroic $BiFeO_3$ were studied in [1, 6-9]. We will consider the situation when ferroelectric stripe domain structure with $P_z(x)$ exists in a film; $z$ is taken to be parallel to the normal to a film. The domain sizes in which polarization $P_z$ attains typical values change from several nanometers up to 100 nm while domain wall width is of the order 1 nm [1, 2].

To describe the magnetic moment distribution we turn to a free energy of a system,

$$f = f_{exch} + f_{an} + f_{me} \quad (1)$$

where

$$f_{exch} = A\{(\nabla\theta)^2 + \sin^2\theta(\nabla\varphi)^2\}$$

is the exchange energy, $A$ is the stiffness constant, $\theta$, $\varphi$ are the polar and azimuthal angles of antiferromagnetic vector spherical coordinate system with the polar axis aligned with the axis $z$.

$l_x = \sin\theta\cos\varphi$; $l_y = \sin\theta\sin\varphi$; $l_z = \cos\theta$ in the spherical coordinate system with the polar axis aligned with the axis $z$.

$$f_{an} = -K_1\sin^2\theta + K_2\sin^2\theta\cos^2\varphi + \frac{\chi_\perp D_1^2}{2}P_z^2\sin^2\theta$$

is energy of magnetic anisotropy including also contribution due to magnetoelectric Dzyaloshinskii – Moria interaction. We can take into consideration the orthorhombic anisotropy for the model description. Let's the $z$-axis is parallel with the $P$ and $x$-axis is parallel with easy axis in the plane that is perpendicular to the $P$ [11, 18]. Herein $K_1$, $K_2$ are magnetic anisotropy constants, $\chi_\perp$ is magnetic susceptibility, $D_1$ is the constant of the magnetoelectric Dzyaloshinskii – Moria interaction. Thus, account of Dzyaloshinskii – Moria contribution renormalizes the constant of magnetic anisotropy.

The term

$$f_{me} = D_2 P_z \left\{ \cos\varphi \frac{\partial\theta}{\partial x} + \sin\varphi \frac{\partial\theta}{\partial y} \right\}$$

is the inhomogeneous magnetoelectric interaction of the Lifshitz type. $D_2$ is the constant of inhomogeneous magnetoelectric interaction [16].

Minimization of the free-energy functional $F = \int f dV$ by the Lagrange–Euler method gives us the solutions 1) $\varphi=0, \pi$, $\theta=\theta(x)$; 2) $\theta=\pi/2$, $\varphi=\varphi(x)$. That means the two possibilities of spins rotation over domain wall exist. Spins can rotate in a film plane and in a plane oriented along normal to a film. It can be shown that in the first case when $E = 4\sqrt{AK_{eff}} - \pi D_2 P > 0$, where $K_{eff} = \left| -K_1 + K_2 + \frac{\chi_\perp}{2}D_1^2 P_z^2 \right|$, the energy of spins rotating in the plane normal to a surface can be less than the energy of spins rotating in the plane of a film. We can always choose the range of physical parameters of $BiFeO_3$ satisfying to the above – mentioned condition, so we limit ourselves to consider the situation $\varphi=0$, $\theta=\theta(x)$. The Euler – Lagrange equation for (1) is of a form

$$A\frac{d^2\theta}{dx^2} + K_{eff}\sin\theta\cos\theta = D_2\frac{dP_z}{dx} \quad (2)$$

At first, we shall study the case $d \ll 1$, $\theta \ll 1$, where $d$ is the ferroelectric domains width. Ferroelectric domain walls are much thinner than antiferromagnetic ones which is valid for $BiFeO_3$ [1, 19]. Then we suppose that the distribution of polarization vector over the structure is defined by the law

$$\frac{dP_z}{dx} = 2P_0\sum_{n=-\infty}^{\infty}(-1)^n\delta(x-dn) \quad (3)$$

Using the Green functions approach we find a solution of linearized equation (2) and present it in a form

$$\theta(x) = q_0\Delta\left\{ \text{ch}\frac{x}{\Delta}\frac{\exp\left(-\frac{d}{\Delta}\right)}{1+\exp\left(-\frac{d}{\Delta}\right)} - \frac{\exp\left(-\frac{|x|}{\Delta}\right)}{2} \right\} \quad (4)$$

where

$$\Delta = \sqrt{\frac{A}{K_{eff}}}, \quad q_0 = \frac{D_2 P_0}{2A}$$

The function (4) is symmetrical, it attains maximum $\theta=\theta_0$ in the points $x=\pm(2k+1)d$ and minimum $\theta=-\theta_0$ in the points $x=\pm 2kd$, $k=0,1,2,...$, $\theta_0 = \frac{1-\exp(-d/\Delta)}{2(1+\exp(-d/\Delta))}$. Po-larization changes in the same way, it gets negative values $P=-P_0$ at $x \in [2kd, (2k+1)d]$ and positive values $P=P_0$ at $x \in [(2k+1)d, 2kd]$. Distribution of the antiferromagnetic vector is to be periodical in a plane containing polarization vector (*ZOX*).

Thus, at first, we can conclude that in a presence of the ferroelectric stripe domain structure the flexomagnetoelectric mechanism forces the formation of periodical antiferromagnetic structure.

To further understand the role of the flexomagnetoelectric mechanism in a stabilization process of antiferromagnetic domain structure let us consider the dependence of domain structure on the width of ferroelectric domains *d*. In this case we can not apply linearization procedure; equation (2) should be solved directly. We will transform right – hand side of Eq. (2) together with Eq. (3) into boundary conditions which acquire a form

$$\frac{d\theta}{dx} = \pm q_0, \text{ at } x = \frac{d}{2}, \frac{3d}{2}, \frac{5d}{2}, \ldots \quad (5a)$$

$$\theta = \frac{\pi}{2} \text{ in the middle of domain } (x=0, d, 2d, \ldots) \quad (5b)$$

The solution of a (2) is of a form

$$\theta = \arcsin\left(\operatorname{sn}(\pm \frac{x-x_0}{\Delta m}, m)\right) \quad (6)$$

The values of the *m* and the $x_0$ constants are determined from boundary conditions (5a), (5b):

$$x_0 = x_c \mp \Delta m K(m) \quad (7)$$

where $x_c = 0, d, 2d, \ldots$

$$\operatorname{dn}(\pm \frac{x-x_c}{\Delta} + K(m), m) = \pm \Delta q_0 \quad (8)$$

where $\operatorname{sn}(x, m)$, $\operatorname{dn}(x, m)$ are the Jacobi elliptic functions, $K(m)$ is quarter of a period of elliptic function.

Then we can construct dependences $\theta(x)$ using equations (6) – (8) e.g. at the following values of physical parameters: $A=3 \cdot 10^{-7}$ erg/cm, $K=8 \cdot 10^5$ erg/cm$^3$, $P_0=6 \cdot 10^{-5}$ C/cm$^2$, $\chi_\perp \sim 5 \cdot 10^{-5}$, $D_2 = \frac{4\pi A}{\lambda P_0}$, $\lambda=62$nm [16]. Distribution of the antiferromagnetic *l* vector or a plate is periodic in *x* direction.

To pay attention to the principal features of the solution let us consider the change of angle $\theta$ upon *x* at various values of *d* on the interval [-*d*/2, 3*d*/2] (Fig.1). The dependence $\theta(x)$ is linear for the small values of ferroelectric domain width *d* (Fig.1a), but when the *d* increases curve $\theta(x)$ acquires nonlinear form (Fig.1b). Nonlinear behaviour appears mainly close to the boundary of ferroelectric domain for sufficiently large values of *d* (starting from 25 nm). In the insert of Fig.1 the geometry of a problem is presented.

It is seen from Fig.1 that distribution of antiferromagnetic vector is represented by parts of cycloid. Then we can say that ferroelectric domain structure partially restores cycloid.

**Conclusion** Our analysis has shown that in multiferroics possessing with electric polarization of the BiFeO$_3$ type the antiferromagnetic domain structure is coupled with the ferroelectric one. The major mechanism determining the antiferromagnetic pattern and the coupling of the antiferromagnetic and ferroelectric domain structures is the inhomogeneous magnetoelectric interaction of the Lifshitz type interaction. The character of spin arrangement is determined by the boundary conditions and strongly depends on the width of ferroelectric domains. Antiferromagnetic magnetic structure is pinned by the ferroelectric one. The main qualitative feature and the driving force of the pinning is the inhomogeneous magnetoelectric interaction of the Lifshitz type at the boundary of ferroelectric domains.

This work was partially supported by the "Progetto Lagrange-Fondazione CRT" и RFBR (08-02-90060-Bel, 08-02-01068 ).

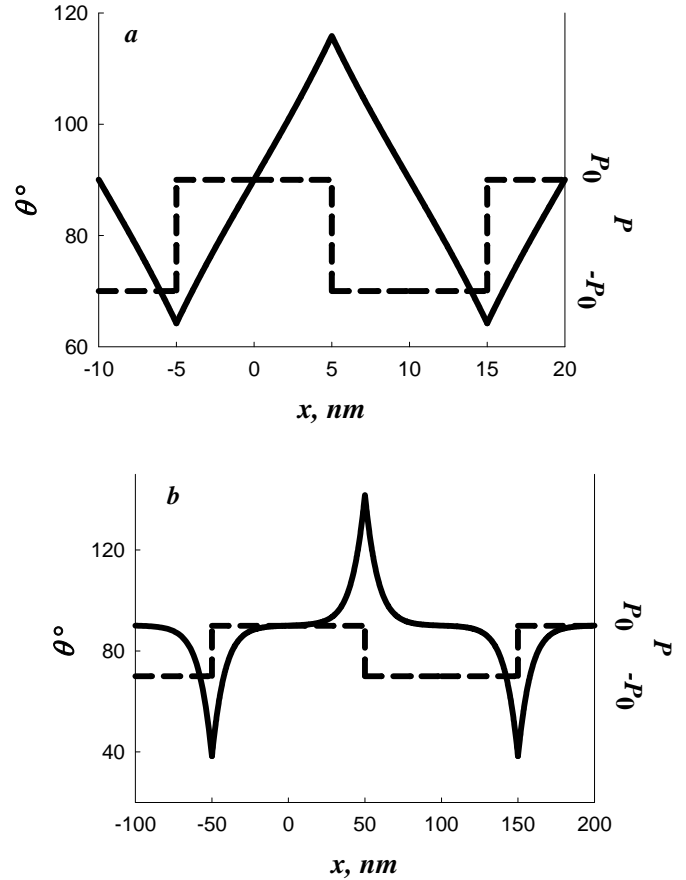

**Figure 1** Dependences of the angle characterizing antiferromagnetic vector distribution $\theta$ on the *x* coordinate, **a)** *d*=10 nm, **b)** *d*=100 nm